\documentclass{JHEP3}

  \usepackage{latexsym,bm,amsmath,amssymb,amsfonts}
  \usepackage{epsfig,graphics,graphicx}
  \usepackage{slashed}
  \usepackage[latin1]{inputenc}
  \usepackage{mathrsfs}
\usepackage{mathcomp}

  \long\def\comment#1{ }

\def\be{\begin{equation}}
\def\ee{\end{equation}}
\def\bea{\begin{eqnarray}}
\def\eea{\end{eqnarray}}

\title{\rm \LARGE \bf Quantum corrections and black hole spectroscopy}

\author{Qing-Quan Jiang$^{a,b}$, Yan Han$^{c}$ and Xu Cai$^{b}$\\

$^{a}$ Institute of Theoretical Physics, China
 West Normal University, Nanchong, Sichuan \\637002, People's Republic of
China\\
$^{b}$ Institute of Particle Physics, Central China Normal
University, Wuhan, Hubei 430079, \\People's Republic of China\\
$^{c}$ College of Mathematic and Information, China
 West Normal University, Nanchong, \\Sichuan 637002, People's Republic of China\\
{\tt E-mail address: jiangqq@iopp.ccnu.edu.cn, xcai@mail.ccnu.edu.cn}}

\abstract{In the work \cite{BRM,RBE}, black hole spectroscopy has been successfully reproduced in the tunneling picture. As a result, the derived entropy spectrum of black hole in different gravity (including Einstein's gravity, Einstein-Gauss-Bonnet gravity and Ho\v{r}ava-Lifshitz gravity) are all evenly spaced, sharing the same forms as $S_n=n$, where physical process is only confined in the semiclassical framework. However, the real physical picture should go beyond the semiclassical approximation. In this case, the physical quantities would undergo higher-order quantum corrections, whose effect on different gravity shares in different forms. Motivated by these facts, in this paper we aim to observe how quantum corrections affect black hole spectroscopy in different gravity. The result shows that, in the presence of higher-order quantum corrections, black hole spectroscopy in different gravity still shares the same form as $S_n=n$, further confirming the entropy quantum is universal in the sense that it is not only independent of black hole parameters, but also independent of higher-order quantum corrections. This is a desiring result for the forthcoming quantum gravity theory.}


\begin{document}

\section{Introduction}
Many proposals have been successfully applied to present Hawking radiation from black hole since Hawking discovered that black hole is not completely black, but can radiate particles with a perfect blackbody spectrum \cite{r1}. Among them, the tunneling method has been widely discussed, because it reflects the real process of particle emission from black hole horizon \cite{r3,MF,t1,t2,t3}. In these observations, the semiclassical Hawking temperature of black hole could be correctly reproduced in the tunneling framework, however, they failed to yield directly the blackbody spectrum. To rectify this shortcomings, Banerjee and Majhi, with the aid of density matrix techniques, reformulated the tunneling mechanism to directly find the spectrum \cite{c1,c2}. Then, based on the modified version of the tunneling mechanism, the entropy spectrum of black hole in different gravity (including Einstein's gravity, Einstein-Gauss-Bonnet gravity and Ho\v{r}ava-Lifshitz gravity) are also described in the tunneling picture \cite{BRM,RBE}. The result shows that, the entropy spectrum in Einstein gravity, Einstein-Gauss-Bonnet gravity as well as Ho\v{r}ava-Lifshitz gravity, are all evenly spaced by the same form as $S_{\textrm{bh}}=n$. In other words, the entropy quantum is independent of black hole parameters in this case.

On the other hand, the entropy spectrum in the tunneling picture \cite{BRM,RBE} is only described in the semiclassical framework. However, the true tunneling process should go beyond the semiclassical approximation. In this case, the physical quantities would undergo higher-order quantum corrections.
Considering full quantum corrections, Banerjee and Majhi first produced higher-order corrections to the thermodynamic entities of black hole \cite{t7}. Later on, quantum corrections to the semiclassical tunneling picture were widely developed in different gravity, such as
Einstein gravity, Einstein-Gauss-Bonnet gravity and Ho\v{r}ava-Lifshitz gravity, etc. As a result, such quantum corrections share in different forms for black hole in different gravity. Now, a question arise whether the modified entropy would cause corrections to the entropy spectrum. For quantization procedure, the answer is expected to be ``Yes" since, in this case, the entropy quantum is universal in the sense that it should be not only independent of black hole parameters, but also independent of quantum corrections.

In this paper, beyond the semiclassical approximation, we first produce black hole entropy in different gravity to include higher-order quantum corrections in the tunneling picture, the result showing that, for black hole in different gravity, higher-order quantum corrections are sharing different forms. For black hole in Einstein gravity, the leading quantum correction to black hole entropy is logarithmic, and next to leading correction is inverse of horizon area. For black hole in Einstein-Gauss-Bonnet and Ho\v{r}ava-Lifshitz gravity, the above logarithmic and inverse corrections are no longer hold, but still contain higher-order corrections. Then, considering the effect of higher-order quantum corrections, we revisit the entropy spectrum in the tunneling picture. As a result, the entropy quantum is independent of the effect of quantum corrections. For black hole in different gravity (including Einstein gravity, Einstein-Gauss-Bonnet gravity and Ho\v{r}ava-Lifshitz gravity), although quantum corrections to black hole entropy take different forms, the entropy spectroscopy share the same forms as $S_n=n$. This is an interesting result since it is the basic requirements of the forthcoming quantum gravity theory.

This paper is organized as follows: In Sec.\ref{sec1}, we briefly review the modified version of the tunneling method beyond the semiclassical approximation, and, with the aid of the first law of black hole thermodynamics, obtain quantum corrections to the black hole entropy in different gravity. The entropy spectroscopy in different gravity is produced in Sec.\ref{sec2} to include higher-order quantum corrections. Sec.\ref{sec3} contains conclusion and discussion.

\section{Quantum corrections and black hole thermodynamics} \label{sec1}

Hawking radiation can be described in the tunneling picture. In the semiclassical approximation, the action for the process of s-wave emission across the horizon
can be reproduced in the Hamilton-Jacobi method, which in turn, is related to the Boltzmann factor for the emission at the Hawking temperature. However, the real physical process should go beyond the semiclassical approximation, in this case, the thermodynamic quantities (including Hawking temperature and black hole entropy) would undergo higher-order quantum corrections. For black hole in different gravity, such corrections share different forms. In this section, our motivation is to present black hole thermodynamics in different gravity to include higher-order quantum corrections.

\subsection{Quantum tunneling beyond the semiclassical approximation}

In this subsection, by considering all the terms in the expansion of the one particle action, we formulate the Hamilton-Jacobi method beyond the semiclassical approximation. For black hole in different gravity, we only focus on the case of the spherically symmetric static solution without loss of generality. For other solutions, near the horizon, the original higher-dimensional theory can always be reduced to the effective two-dimensional theory with gauge potential with respect to the electric gauge symmetry or/and the rotating symmetry. So, it is, without loss of generality, to confine our attention on the case of the spherically symmetric static solution. The generally spherically symmetric black hole takes the form as
\begin{equation}
ds^2=-\mathcal{F}(r)dt^2+\frac{dr^2}{\mathcal{G}(r)}+r^2d\Omega_{(n-2)}^2. \label{eq1}
\end{equation}
For black hole in different gravity, the metric (\ref{eq1}) shares the same form, only $\mathcal{F}(r)$ and $\mathcal{G}(r)$ taking different forms. For example:

\textbf{Case I:} When
\begin{equation}
\mathcal{F}(r)=\mathcal{G}(r)=1-\Big(\frac{r_h}{r}\Big)^{n-3}, \label{neq2.2}
\end{equation}
the metric (\ref{eq1}) describes the Schwarzschild black hole in Einstein gravity. Here, $r_h^{n-3}=\frac{16\pi M}{(n-2)A_{(n-2)}}$, with $A_{(n-2)}$ and $M$, respectively, denoting the volume of a unit $(n-2)$-sphere $d\Omega_{(n-2)}^2$ and the ADM mass of the black hole.

\textbf{Case II:} When
\begin{equation}
\mathcal{F}(r)=\mathcal{G}(r)=1+\frac{r^2}{\widetilde{\lambda}}\Big[1+\epsilon\big(1+\frac{2\omega\widetilde{\lambda}}{r^{n-1}}\big)^{1/2}\Big],\label{eq2.2}
\end{equation}
with $\widetilde{\lambda}=\lambda(n-3)(n-4)$, the metric (\ref{eq1}) describes the spherically symmetric static black hole in Einstein-Gauss-Bonnet gravity \cite{wm}.  Here, the constant $\epsilon$ has the values of $\pm 1$. In this paper, we take the value of $\epsilon=-1$, corresponding to an asymptotically flat metric for (\ref{eq1}). In this case, $\omega=\frac{16\pi M}{(n-2)A_{(n-2)}}$, and the parameters $\lambda$, $A_{(n-2)}$ and $m$ are, respectively, denoting the coupling constant of the Gauss-Bonnet term, the volume of unit $(n-2)$ sphere and the ADM mass of the black hole.

\textbf{Case III:} When
\begin{eqnarray}
\mathcal{G}(r)&=&a+x^2-\alpha x^{(2\lambda\pm \sqrt{6\lambda-2})/(\lambda-1)},\nonumber\\
\mathcal{F}(r)&=& x^{-2(1+3\lambda\pm 2\sqrt{6\lambda-2})/(\lambda-1)}\mathcal{G}(r),\label{eq2.3}
\end{eqnarray}
the metric (\ref{eq1}) is the spherically symmetric static black hole solution at the Lifshitz point $z=3$ in Ho\v{r}ava-Lifshitz theory \cite{ncco}. Here, $a=n-2$ and $d\Omega_a^2$ denotes the line element for a two dimensional Einstein space with constant scalar curvature $2a$. Without loss of generality, $a$ can take the values of $0,\pm 1$, respectively. Also, $x=r\sqrt{-\Lambda}$ and $\Lambda$ is the cosmological constant. In this paper, we take the coupling constant $\lambda=1$, the negative sign for (\ref{eq2.3}) and the negative cosmological constant $\Lambda$. In the limit $\lambda\rightarrow 1$, Eq.(\ref{eq2.3}) could be transformed as
\begin{equation}
\mathcal{F}(r)=\mathcal{G}(r)=a+x^2-\alpha\sqrt{x}.\label{eq2.4}
\end{equation}
In this case, the metric (\ref{eq1}) is asymptotically $AdS_4$, having a singularity at $x=0$ if $\alpha\neq 0$, which could be covered by black hole horizon at $x_h$, the largest root of $f(x_h)=0$.

Now, we focus on studying the tunneling character beyond the semiclassical approximation for black hole in the above different cases. We start our analysis by considering the massless scalar field governed by the Klein-Gordon equation $-\frac{1}{\sqrt{-g}}\partial_\mu\big[g^{\mu\nu}\sqrt{-g}\partial_\nu\big]\Phi=0$.
Since, in the tunneling picture, particles radiate radially, it is enough to consider the ($r-t$) sector of (\ref{eq1}) for the tunneling process. However, in this case, the Klein-Gordon equation can not be solved exactly. To exactly be solved, we choose the standard ansatz for $\Phi$, which is
\begin{equation}
\Phi=\exp{\big[-\frac{i}{\hbar}\mathcal{S}(r,t)\big]},\label{eq2.5}
\end{equation}
where
\begin{equation}
\mathcal{S}(r,t) =\mathcal{S}_0(r,t)+\sum_{i=1}^\infty \hbar^i\mathcal{S}_i(r,t),
\end{equation}
contains higher-order quantum corrections. Substituting Eq.(\ref{eq2.5}) into the Klein-Gordon equation, and equating the coefficients of different orders in $\hbar$ to zero, one can, interestingly, find that each partial differential equation about $\mathcal{S}_n(r,t)$ (n=0,i; i=1,2,3,.....) shares the same form as
\begin{equation}
\frac{\partial \mathcal{S}_n(r,t)}{\partial t}=\pm \sqrt{\mathcal{F}(r)\mathcal{G}(r)}\frac{\partial \mathcal{S}_n(r,t)}{\partial r}.\label{eq2.7}
\end{equation}
So the solutions of $\mathcal{S}_n$ are not independent, and $\mathcal{S}_i$'s are proportional to $\mathcal{S}_0$ with a proportionality factor. After a dimensional analysis\footnote{The units of Newton's constant and mass density are same in all dimensions, which means $[G]\frac{M}{L^3}=[G^n]\frac{M}{L^{n-1}}$. Here, $G$ is the Newton's constant in four dimensions, taking the form as $[G]=\frac{[c]^3L^2}{[\hbar]}$. Combining with these dimensional analysis and replacing $L$ with $n-$dimensional Plank length ($l_p^n$), we obtain $(l_p^n)^{n-2}=\frac{[\hbar]G^n}{[c]^3}$. In this case, when $G^n=c=1$, we have $[\hbar]=(l_p^n)^{n-2}$. For black hole (\ref{eq1}), the only parameter having the unit of length is the horizon radius $r_h$. So, $[\hbar]$ has the same dimension of $r_h^{n-2}$ \cite{t9}. }, the most general form of $\mathcal{S}(r,t)$ can be written as
\begin{equation}
\mathcal{S}(r,t)=\Big(1+\sum_{i=1}^\infty\beta_i\frac{\hbar^i}{(r_h^{n-2})^i}\Big)\mathcal{S}_0(r,t),\label{eq2.8}
\end{equation}
where $\beta_i$ are dimensionless constants, and $r_h$ is the horizon radius for black hole in different gravity. Also, $\mathcal{S}_0(r,t)$ satisfies Eq.(\ref{eq2.7}), so $\mathcal{S}(r,t)$ can be rewritten as
\begin{equation}
\mathcal{S}(r,t)=\Big(1+\sum_{i=1}^\infty\beta_i\frac{\hbar^i}{(r_h^{n-2})^i}\Big)\Big(\omega t\pm \omega \int \frac{dr}{\sqrt{\mathcal{F}(r)\mathcal{G}(r)}}\Big).\label{eq2.9}
\end{equation}
Here, $\omega$ is the conserved quantity with respect to the time-like Killing vector field. If, letting $r_*= \int \frac{dr}{\sqrt{\mathcal{F}(r)\mathcal{G}(r)}}$, the solution for the scalar field to include higher-order quantum corrections takes the form as
\begin{equation}
\Phi=\exp\Big[-\frac{i}{\hbar}\omega\Big(1+\sum_{i=1}^\infty\beta_i\frac{\hbar^i}{(r_h^{n-2})^i}\Big)(t\pm r_*)\Big].
\end{equation}
Now, the right and left modes defined inside and outside the horizon can be written as
\begin{eqnarray*}
\Phi_{\textrm{in}}^{(R)}&=&\exp\Big[-\frac{i}{\hbar}\omega\Big(1+\sum_{i=1}^\infty\beta_i\frac{\hbar^i}{(r_h^{n-2})^i}\Big)u_{\textrm{in}}\Big];\nonumber\\
\Phi_{\textrm{in}}^{(L)}&=&\exp\Big[-\frac{i}{\hbar}\omega\Big(1+\sum_{i=1}^\infty\beta_i\frac{\hbar^i}{(r_h^{n-2})^i}\Big)v_{\textrm{in}}\Big];\nonumber\\
\end{eqnarray*}
\begin{eqnarray}
\Phi_{\textrm{out}}^{(R)}&=&\exp\Big[-\frac{i}{\hbar}\omega\Big(1+\sum_{i=1}^\infty\beta_i\frac{\hbar^i}{(r_h^{n-2})^i}\Big)u_{\textrm{out}}\Big];\nonumber\\
\Phi_{\textrm{out}}^{(L)}&=&\exp\Big[-\frac{i}{\hbar}\omega\Big(1+\sum_{i=1}^\infty\beta_i\frac{\hbar^i}{(r_h^{n-2})^i}\Big)v_{\textrm{out}}\Big].
\end{eqnarray}
Here, $v=t-r_{*}$ and $u=t+r_*$ denote the advanced and retarded coordinates, respectively. In the modified version of the tunneling picture, the Kruskal-like coordinate system is applied to describe the metric (\ref{eq1}) inside and outside the horizon \cite{BRM,RBE,c1,c2}. In this case, we can readily connect the right and left moving modes defined inside and outside the horizon, which is
\begin{eqnarray}
\Phi_{\textrm{in}}^{(R)}&=&\exp\Big[-\frac{\pi \omega}{\hbar\kappa}\Big(1+\sum_{i=1}^\infty\beta_i\frac{\hbar^i}{(r_h^{n-2})^i}\Big)\Big]
\Phi_{\textrm{out}}^{(R)}; \nonumber\\
\Phi_{\textrm{in}}^{(L)}&=&\Phi_{\textrm{out}}^{(L)},\label{eq2.12}
\end{eqnarray}
where $\kappa$ is the surface gravity for black hole in different gravity without presence of higher-order quantum corrections. Now, basing on the relation  (\ref{eq2.12}) between the right and left modes defined inside and outside the horizon, we focus on studying the modes' tunneling across the horizon.
For the left modes, the observer staying outside the horizon would find that it can classically fall into the center of black hole. So the tunneling rate for the left modes should be expected to be unity. This can be easily verified by $P^{(L)}=|\Phi_{\textrm{in}}^{(L)}|^2=|\Phi_{\textrm{out}}^{(L)}|^2$=1. Proceeding in the similar way, for the right modes, its probability, as seen by an external observer, is
\begin{eqnarray}
P^{(R)}=|\Phi_{\textrm{in}}^{(R)}|^2&=&\Big|\exp\Big[-\frac{\pi \omega}{\hbar\kappa}\Big(1+\sum_{i=1}^\infty\beta_i\frac{\hbar^i}{(r_h^{n-2})^i}\Big)\Big]
\Phi_{\textrm{out}}^{(R)}\Big|^2\nonumber\\
&=&\exp\Big[-\frac{2\pi \omega}{\hbar\kappa}\Big(1+\sum_{i=1}^\infty\beta_i\frac{\hbar^i}{(r_h^{n-2})^i}\Big)\Big].\label{eq2.13}
\end{eqnarray}
The total tunneling rate is then given by $P=\frac{P^{(R)}}{P^{(L)}}=P^{(R)}$. When it is related to the Boltzmann factor for the emission at the Hawking temperature, the Hawking temperature to include higher-order quantum corrections can be reproduced. Then, with the aid of the first law of thermodynamics, we can also present quantum corrections to the black hole entropy. In the following subsection, we focus on studying black hole thermodynamics in different gravity beyond the semiclassical approximation.

\subsection{Black hole thermodynamics}

In this subsection, we aim to produce black hole thermodynamics in different gravity to include higher-order quantum corrections. When the total tunneling rate $\Gamma$ is related to the Boltzmann factor, we can easily find, from Eq.(\ref{eq2.13}), that the Hawking temperature to include higher-order quantum corrections is
\begin{equation}
T_{\textrm{corr}}=T_h\Big(1+\sum_{i=1}^\infty\beta_i\frac{\hbar^i}{(r_h^{n-2})^i}\Big)^{-1},\label{eq2.14}
\end{equation}
where $T_h=\frac{\hbar \kappa}{2\pi}$ is the Hawking temperature without presence of higher-order quantum corrections. Next, with the help of the quantum corrected Hawking temperature (\ref{eq2.14}) and the first law of thermodynamics, we aim to produce black hole thermodynamics beyond the semiclassical approximation for black hole in different gravity:

\textbf{Case I:} For the Schwarzschild black hole in Einstein gravity with the metric (\ref{eq1}) and (\ref{neq2.2}), according to Eq.(\ref{eq2.14}), the Hawking temperature with higher-order quantum corrections is given by
\begin{equation}
T_{\textrm{corr}}^\textrm{S}=\frac{\hbar(n-3)}{4\pi r_h}\Big(1+\sum_{i=1}^\infty\beta_i\frac{\hbar^i}{(r_h^{n-2})^i}\Big)^{-1}.\label{eq2.16}
\end{equation}
With the aid of the corrected Hawking temperature (\ref{eq2.16}), we can proceed with the
calculation of the modified black hole entropy. In the presence of higher-order quantum corrections, we assume that the first
law of thermodynamics of the black hole still complies with $dS_{\textrm{bh}}^\textrm{S}=(T_{\textrm{corr}}^\textrm{S})^{-1}dM$. In this case, the quantum-corrected entropy can be written as
\begin{eqnarray}
S_{\textrm{bh}}^\textrm{S}&=&\frac{1}{4\hbar}A_{(n-2)}r_h^{n-2}+\frac{(n-2)\beta_1A_{(n-2)}}{4}\ln r_h \nonumber\\
&-&\frac{\beta_2\hbar A_{(n-2)}}{4r_h^{n-2}}+\textrm{constant}+\textrm{higher order terms in $\hbar$}.
\label{eq2.17}
\end{eqnarray}
Here, the first term is the semiclassical entropy of the Schwarzschild black hole in Einstein gravity, which is related to one-quarter of its horizon area, also called as the Bekenstein-Hawking entropy. The other term come from higher-order quantum corrections. Obviously, beyond the semiclassical approximation, the semiclassical entropy would undergo higher quantum corrections in the sense that the leading correction is logarithmic and next to the leading correction is inverse of the horizon area.

\textbf{Case II:} Similarly, for the spherically symmetric static black hole in Einstein-Gauss-Bonnet gravity describing in (\ref{eq1}) and (\ref{eq2.2}), beyond the semiclassical approximation, the Hawking temperature is corrected as
\begin{equation}
T_{\textrm{corr}}^{\textrm{GB}}=\frac{\hbar}{8\pi r_h}\Big[\frac{2(n-3)r_h^2+(n-5)\widetilde{\lambda}}{r_h^2+\widetilde{\lambda}}\Big]\Big(1+\sum_{i=1}^\infty\beta_i\frac{\hbar^i}{(r_h^{n-2})^i}\Big)^{-1}.\label{eq2.18}
\end{equation}
In view of the quantum-corrected Hawking temperature (\ref{eq2.18}), the first law of thermodynamic would present the modified entropy taking the form as
\begin{eqnarray}
S_{\textrm{bh}}^{\textrm{GB}}&=&\frac{1}{4\hbar}A_{(n-2)}r_h^{n-2}\Big(1+\frac{(n-2)\widetilde{\lambda}}{(n-4)r_h^2}\Big)+\frac{(n-2)\beta_1A_{(n-2)}}{4}\Big(\ln r_h-\frac{\widetilde{\lambda}}{2r_h^2}\Big) \nonumber\\
&-&\frac{\beta_2\hbar A_{(n-2)}}{4}\Big(\frac{1}{r_h^{n-2}}+\frac{(n-2)\widetilde{\lambda}}{nr_h^n}\Big)+\textrm{constant}+\textrm{higher order terms in $\hbar$}. \label{eq2.19}
\end{eqnarray}
In (\ref{eq2.19}), the first term is the semiclassical entropy of the spherically symmetric static black hole in Einstein-Gauss-Bonnet gravity \cite{nms1}, and the other terms come from the higher-order quantum corrections. Obviously, in the presence of the coupling constant of the Gauss-Bonnet term $\lambda$, the semiclassical entropy is no longer one-quarter of its horizon area, and the leading correction, as a result of quantum corrections, is not purely logarithmic and also next to the leading correction is not inverse of the horizon area. This is in sharp contrast to that for black hole in Einstein gravity, where the entropy is corrected in the form as (\ref{eq2.17}). Only when eliminating the Gauss-Bonnet coupling constant $\lambda$, the entropy can be corrected in the way as that for black hole in the usual Einstein gravity. So, the presence of Gauss-Bonnet term would modify the results in a nontrivial manner.

\textbf{Case III:} For the spherically symmetric static black hole solution at the Lifshitz point $z=3$ in Ho\v{r}ava-Lifshitz theory describing in (\ref{eq1}) and (\ref{eq2.4}), according to Eq.(\ref{eq2.14}), the Hawking temperature to include higher-order quantum corrections is read off
\begin{equation}
T_{\textrm{corr}}^{\textrm{HL}}=\frac{\hbar(3x_h^2-a)}{8\pi x_h}\sqrt{-\Lambda}\Big(1+\sum_{i=1}^\infty\beta_i\frac{\hbar^i}{(x_h^2)^i}\Big)^{-1}.\label{eq2.20}
\end{equation}
With the quantum-corrected Hawking temperature (\ref{eq2.20}), we will obtain the modified black hole entropy by using the first law of black hole thermodynamics
with assumption that as a thermodynamical system, the first law always keeps valid. Integrating this relation yields
\begin{eqnarray}
S_{\textrm{bh}}^{\textrm{HL}}&=&\frac{\pi a^2 \mu^2A_a}{4\hbar}(x_h^2+2a\ln x_h)+\frac{\pi a^2 \mu^2A_a \beta_1}{4}\Big(2\ln x_h-\frac{a}{x_h^2}\Big)\nonumber\\
&-& \frac{\pi a^2 \mu^2A_a\beta_2 \hbar}{8}\Big(\frac{2}{x_h^2}+\frac{a}{x_h^4}\Big)+\textrm{constant}+\textrm{higher order terms in $\hbar$}. \label{eq2.21}
\end{eqnarray}
The first term in (\ref{eq2.21}) is the semiclassical entropy of black hole in Ho\v{r}ava-Lifshitz gravity, the result showing that it contains an additive term proportional to logarithmic of area, apart from the area term as usually happens in Einstein gravity. Moreover, when the higher-order quantum corrections are included, the entropy of black hole in Ho\v{r}ava-Lifshitz gravity is also corrected in a nontrivial way in contrast with that in Einstein gravity. Also, such corrections behave not just like that in Einstein-Gauss-Bonnet gravity.

In a word, for black hole in different gravity, the semiclassical entropy and its higher-order quantum corrections both share different forms, whose presence refers to (\ref{eq2.17}), (\ref{eq2.19}) and (\ref{eq2.21}). On the other hand, the entropy spectrum is directly related to the black hole entropy. Combined with these facts, in the following section, our focus is on finding the entropy spectrum of the different-gravity black hole in the presence of quantum corrections to see how such corrections affect the entropy spectrum.

\section{Quantum corrections and black hole spectroscopy in different gravity} \label{sec2}

Beyond the semiclassical approximation, the thermodynamic entities of black hole would undergo higher-order quantum corrections. In addition, for black hole in different gravity, such corrections share different forms. In view of these development, in this section, our motivation is to check how such quantum corrections affect black hole spectroscopy in different gravity.

\subsection{Black hole spectroscopy in Einstein gravity}

For black hole in Einstein gravity, the thermodynamic entities are corrected in the manner as (\ref{eq2.16}) and (\ref{eq2.17}) beyond the semiclassical approximation. In this case, the right and left modes inside and outside the horizon are defined in (\ref{eq2.12}) only replacing the surface gravity $\kappa$ with that of the Schwarzschild black hole. In the previous section, we have shown that when pair production occurs inside the horizon, the left mode is trapped inside the horizon, while the right mode can tunnel across the horizon to be observed at the asymptotic infinity. Thus, the average value of $\omega$ is given by
\begin{equation}
\langle\omega\rangle=\frac{\int_0^\infty(\Phi_{\textrm{in}}^{(R)})^*\omega\Phi_{\textrm{in}}^{(R)}d\omega}
{\int_0^\infty(\Phi_{\textrm{in}}^{(R)})^*\Phi_{\textrm{in}}^{(R)}d\omega}.\label{eq3.1}
\end{equation}
Here, the average value of the energy $\omega$ is related to the right mode inside the horizon. Actually, the observer is lived outside the horizon, so it is necessary to recast the ``in"  quantity into its ``out" correspondence, which could be well done in (\ref{eq2.12}). Now, we have
\begin{equation}
\langle\omega\rangle=\frac{\int_0^\infty \exp{\big(-\frac{\omega}{T_{\textrm{corr}}^\textrm{S}}\big)}(\Phi_{\textrm{out}}^{(R)})^*\omega\Phi_{\textrm{out}}^{(R)}d\omega}
{\int_0^\infty\exp{\big(-\frac{\omega}{T_{\textrm{corr}}^\textrm{S}}\big)}(\Phi_{\textrm{out}}^{(R)})^*\Phi_{\textrm{out}}^{(R)}d\omega}
=T_{\textrm{corr}}^\textrm{S}.\label{eq3.2}
\end{equation}
In a similar way, the average squared energy of the particle detected by the asymptotic observer is obtained as $\langle\omega^2\rangle={2}{T_{\textrm{corr}}^\textrm{S}}$. Now, we can easily find the uncertainty in the detected energy $\omega$ is
\begin{equation}
\Delta \omega =\sqrt{\langle\omega^2\rangle-\langle\omega\rangle^2}=T_{\textrm{corr}}^\textrm{S}.\label{eq3.3}
\end{equation}
This uncertainty can be treated as the lack of information in energy of the black hole during the particle emission. Also, in the information theory,
the entropy acts as the lack of information. To connect these two quantities, we can use the first law of thermodynamic as $T_{\textrm{corr}}^{\textrm{S}}\Delta S_{\textrm{bh}}^{\textrm{S}}=\Delta \omega$. In view of (\ref{eq3.3}), the uncertainty in the characteristic frequency of the outgoing mode is $\Delta f=\frac{\Delta\omega}{\hbar}=\frac{T_{\textrm{corr}}^\textrm{S}}{\hbar}$. According to the Bohr-Sommerfeld quantization rule, substituting this uncertainty into the first law of thermodynamic yields the entropy spectrum as
\begin{equation}
 S_{\textrm{bh}}^{\textrm{S}}=n,\label{eq3.4}
\end{equation}
where $n$ is an integer. From Eq.(\ref{eq3.4}), $\Delta S_{\textrm{bh}}^{\textrm{S}}=(n+1)-n=1$, this result showing that the entropy of the Schwarzschild black hole is quantized in units of the identity. It should be noted that the result (\ref{eq3.4}) is obtained in the presence of higher-order quantum corrections.
Also, in Ref.\cite{RBE}, the same result was reproduced in the semiclassical tunneling picture. So, we can conclude the entropy quantum is independent of black hole parameters to include higher-order quantum corrections or not.

\subsection{Black hole spectroscopy in Einstein-Gauss-Bonnet gravity}

Similarly, for black hole in Einstein-Gauss-Bonnet gravity, the average value of $\omega$ is related to the right mode outside the horizon by
\begin{equation}
\langle\omega\rangle=\frac{\int_0^\infty \exp{\big(-\frac{\omega}{T_{\textrm{corr}}^\textrm{GB}}\big)}(\Phi_{\textrm{out}}^{(R)})^*\omega\Phi_{\textrm{out}}^{(R)}d\omega}
{\int_0^\infty\exp{\big(-\frac{\omega}{T_{\textrm{corr}}^\textrm{GB}}\big)}(\Phi_{\textrm{out}}^{(R)})^*\Phi_{\textrm{out}}^{(R)}d\omega}
=T_{\textrm{corr}}^\textrm{GB}.\label{eq3.5} 
\end{equation}
Also, the average squared energy of the particle detected by the asymptotic observer can be derived as $\langle\omega^2\rangle={2}{T_{\textrm{corr}}^\textrm{GB}}$. So, the uncertainty in the detected energy $\omega$ is given by
\begin{equation}
\Delta \omega =\sqrt{\langle\omega^2\rangle-\langle\omega\rangle^2}=T_{\textrm{corr}}^\textrm{GB}.\label{eq3.6}
\end{equation}
In a similar way, with the aid of the first law of thermodynamic and the uncertainty in the characteristic frequency of the outgoing mode $\Delta f=\frac{T_{\textrm{corr}}^\textrm{GB}}{\hbar}$, the
Bohr-Sommerfeld quantization rule yields the entropy spectrum taking the form as
\begin{equation}
 S_{\textrm{bh}}^{\textrm{GB}}=n,\label{eq3.7}
\end{equation}
where $n$ is an integer. From (\ref{eq3.7}), we can easily find $\Delta S_{\textrm{bh}}^{\textrm{GB}}=1$, showing that the entropy of black hole in Einstein-Gauss-Bonnet gravity is still quantized in units of the identity even in the presence of higher-order quantum corrections. In Ref.\cite{RBE}, the same result is semiclassically reproduced in the tunneling framework. As a conclusion, the entropy quantum is independent of higher-order quantum corrections.

\subsection{Black hole spectroscopy in Ho\v{r}ava-Lifshitz gravity}

For black hole in Ho\v{r}ava-Lifshitz gravity, we can similarly reproduce the average value of $\omega$ as
\begin{equation}
\langle\omega\rangle=\frac{\int_0^\infty \exp{\big(-\frac{\omega}{T_{\textrm{corr}}^\textrm{HL}}\big)}(\Phi_{\textrm{out}}^{(R)})^*\omega\Phi_{\textrm{out}}^{(R)}d\omega}
{\int_0^\infty\exp{\big(-\frac{\omega}{T_{\textrm{corr}}^\textrm{HL}}\big)}(\Phi_{\textrm{out}}^{(R)})^*\Phi_{\textrm{out}}^{(R)}d\omega}
=T_{\textrm{corr}}^\textrm{HL}.\label{eq3.8}
\end{equation}
Also, the average squared energy of the particle detected by the asymptotic observer can be obtained as $\langle\omega^2\rangle={2}{T_{\textrm{corr}}^\textrm{HL}}$. Combined with these facts, the uncertainty in the detected energy $\omega$ is written as
\begin{equation}
\Delta \omega =\sqrt{\langle\omega^2\rangle-\langle\omega\rangle^2}=T_{\textrm{corr}}^\textrm{HL}.\label{eq3.9}
\end{equation}
In a similar way, the entropy spectrum of the black hole in Ho\v{r}ava-Lifshitz gravity appears in the presence of higher-order quantum corrections as
\begin{equation}
 S_{\textrm{bh}}^{\textrm{HL}}=n.\label{eq3.10}
\end{equation}
Obviously, in the presence of higher-order quantum corrections, the entropy of black hole in Ho\v{r}ava-Lifshitz gravity is still quantized
in units of the identity, just like that in the semiclassical case \cite{BRM}. So, for black hole in Ho\v{r}ava-Lifshitz gravity, the entropy quantum is still
independent of higher-order quantum corrections.

As a summary of this section, the entropy quantum is universal in the sense that it is not only independent of black hole parameters, but also independent of higher-order quantum corrections. For black hole in different gravity, although the quantum-corrected entropy shares different forms, the entropy quantum share the same forms as $S_{\textrm{bh}}=n$.

\section{Conclusion and Discussion}\label{sec3}

In this paper, beyond the semiclassical approximation, we first produce black hole entropy in different gravity to include higher-order quantum corrections in the tunneling picture, the result showing that, for black hole in different gravity, the semiclassical entropy accompanying with its higher-order quantum corrections
are both sharing different forms. In detail, for black hole in Einstein gravity, the semiclassical entropy obey the Bekenstein-Hawking area law, and its leading correction is logarithmic and next to the leading correction is inverse of horizon area. For black hole in Einstein-Gauss-Bonnet gravity, due to the presence of Gauss-Bonnet coupling terms, from Eq.(\ref{eq2.19}) we would find different result from that in Einstein gravity. Here, the semiclassical entropy is not just one-quarter of its horizon area, containing an additive term which can be written as a function of horizon area. Also, the leading correction is not pure logarithmic and next to the leading correction is not inverse of the horizon area. Additionally, the similar result can be find in (\ref{eq2.21}) for black hole
in Ho\v{r}ava-Lifshitz gravity with higher-order quantum corrections, where the semiclassical entropy contains an additive term proportional to logarithmic of area, and higher-order quantum corrections also behave in a nontrivial way in contrast with that in Einstein and Einstein-Gauss-Bonnet gravity. Then, in the presence of higher-order quantum corrections, we produce
the entropy spectrum of black hole in different gravity. The result shows that, although the semiclassical entropy accompanying with its higher-order quantum corrections are both sharing different forms for black hole in different gravity, the entropy spectrum shares the same form as $S_{\textrm{bh}}=n$, which shows that the entropy quantum is not only independent of black hole parameters, but also independent of higher-order quantum corrections. This is a desiring result for the forthcoming quantum gravity theory.

In this paper, quantum concept is reflected in the entropy spectrum, rather than the area spectrum. For black hole in Einstein gravity, the semiclassical entropy is related to the horizon area by the Bekenstein-Hawking formula $S_{\textrm{bh}}={A}/[{4(l_{p}^n)^{n-2}}]$, where $l_{p}^n$ is the $n$-dimensional Planck length. In this case, when the entropy spectrum is $S_{\textrm{bh}}=n$, the area spectrum is evenly spaced as $A_n=4(l_{p}^n)^{n-2} n$. Obviously, the area quantum $\Delta A_n=4(l_{p}^n)^{n-2}$ is in agreement with that of Hod by considering the Heisenberg uncertainty principle and Schwinger-type charge emission process \cite{hod}, but is smaller than that of Bekenstein \cite{bek} and that from the quasinormal modes by Hod and Kunstatter \cite{qmo} as well as that from the new interpretation of quasinormal modes given by Maggiore \cite{MVM1} and exploited by Vagenas and Medved \cite{MVM2}. This equality is due to the similarity between the tunneling mechanism and the Schwinger mechanism. However, in our case, black hole entropy contains higher-order quantum corrections, and the entropy spectrum shares the same form as that in the semiclassical case. So, for black hole in Einstein gravity, the entropy spectrum is equispaced in the presence of higher-order quantum corrections, but the corresponding area spectrum is not. On the other hand, for black hole in Einstein-Gauss-Bonnet gravity and Ho\v{r}ava-Lifshitz gravity, we can easily find, from
(\ref{eq2.19}), (\ref{eq2.21}), (\ref{eq3.7}) and (\ref{eq3.10}), the area spectrum is not equispaced in the presence of higher-order quantum corrections, but the entropy has an equispaced spectrum. So, the entropy quantum is more \emph{natural} than the area quantum in our case.

In this paper, we assume that pair production occurs just inside the horizon. According to this scenario, the left modes is
trapped inside the black hole, while the right mode can tunnel through the horizon to be observed at asymptotic infinity. In fact,
this tunneling picture can be depicted in another manner, that is, particles are created just outside the horizon, the
left modes tunnel into the horizon, while the right modes are left outside and emerge at
infinity. In this case, the left modes are tunneling across the horizon, whose appearance should reproduce the same result as that in this paper.

\section*{Acknowledgments}
This work is supported by the Natural Science Foundation of China with Grant Nos.
10773008, 10975062, 70571027, 10635020, a grant by the Ministry of Education of China
under Grant No. 306022.

\end{document}